# Vanadium spin qubits as telecom quantum emitters in silicon carbide

Gary Wolfowicz[1,2], Christopher P. Anderson[1,3], Berk Diler[1], Oleg G. Poluektov[4], F. Joseph Heremans[1,2] and David D. Awschalom[1,2,3,*]

[1]Pritzker School of Molecular Engineering, University of Chicago, Chicago, Illinois 60637, USA
[2]Center for Molecular Engineering and Materials Science Division, Argonne National Laboratory, Lemont, Illinois 60439, USA
[3]Department of Physics, University of Chicago, Chicago, Illinois 60637, USA
[4]Chemical Sciences and Engineering Division, Argonne National Laboratory, Lemont, Illinois 60439, USA
*email: awsch@uchicago.edu

Solid state quantum emitters with spin registers are promising platforms for quantum communication, yet few emit in the narrow telecom band necessary for low-loss fiber networks. Here we create and isolate near-surface single vanadium dopants in silicon carbide (SiC) with stable and narrow emission in the O-band (1278-1388 nm), with brightness allowing cavity-free detection in a wafer-scale CMOS-compatible material. In vanadium ensembles, we characterize the complex $d^1$ orbital physics in all five available sites in 4H-SiC and 6H-SiC. The optical transitions are sensitive to mass shifts from local silicon and carbon isotopes, enabling optically resolved nuclear spin registers. Optically detected magnetic resonance in the ground and excited orbital states reveals a variety of hyperfine interactions with the vanadium nuclear spin and clock transitions for quantum memories. Finally, we demonstrate coherent quantum control of the spin state. These results provide a path for telecom emitters in the solid-state for quantum applications.

**Introduction**

Quantum networks are rapidly emerging with key demonstrations realized using fiber optics and satellites (*1*, *2*). One of the fundamental challenges to practical quantum communication remains the ability to create efficient and scalable quantum repeaters that operate in the telecom range for low loss fiber transmission. Solid-state solutions have involved advances in engineering, including frequency conversion and optical cavities (*3–5*), and the search for better optically active impurities with long spin memories (*6–10*). Current systems include vacancy-related defects for bright single photon emission or rare-earth ions for telecom emission (*4*, *11*), yet few candidates possess both properties.

Transition metal ions such as chromium in ruby have long been used in laser systems for their 3$d$ shell optical transitions, often with emission in the near-infrared (*12*, *13*). Recent works using chromium, molybdenum and vanadium in silicon carbide (SiC) have shown that they may be equally relevant for quantum applications, combining both brightness and (near-)telecom emission in a technologically mature material (*14–16*). Among these ions, vanadium $V^{4+}$ is the only fully telecom transition metal emitter, covering the entire O-band spectrum (1278-1388 nm) depending on its substitutional silicon site in the 4H and 6H polytypes of SiC (*17*, *18*). Though V has been studied for decades for compensating SiC crystals, critical questions remain regarding properties for quantum applications.

In this work, we systematically investigate the optical and spin properties of $V^{4+}$ dopants in all five inequivalent sites of 4H-SiC and 6H-SiC. We present observations of single V emitters implanted into commercial wafers and confirm that their optical properties are stable and reproducible. In ensembles, we characterize the orbital structure through resonant excitation, and observe a strong effect from the nearest-neighbor isotope environment ($^{28}$Si, $^{29}$Si, $^{30}$Si, $^{12}$C and $^{13}$C) of the V dopant. We also obtain lifetime, Debye-Waller factor and hole burning measurements to validate this system as an optical interface. We further present optically detected magnetic resonance (ODMR) and obtain the anisotropic g-factors and hyperfine spin parameters for three of the orbital states, depending on the sites, revealing complex spin Hamiltonians with clock transitions (*19*). Finally, we demonstrate a proof-of-concept coherent driving of the spin state. While some of the experiments are only shown on select sites for conciseness, the measured parameters are summarized in Table 1 for all sites with corresponding data available in the Supplementary Materials. Overall, these results reveal vanadium ions in SiC as prime candidates for telecom quantum networks.

## Results
### Optical spectroscopy of defect ensembles

Vanadium ($^{51}$V isotope, ~100% abundance) sits in the silicon site of SiC as a substitutional dopant (Fig. 1A). It is stable in different charge states including the neutral charge state $V^{4+}$ emitting around 1.3 µm and the negative charge state $V^{3+}$ emitting around 2 µm (*20*). The ground state of $V^{4+}$ is located 1.6 eV above the valence band, about mid-level within the 3-3.2 eV bandgap of 4H-SiC and 6H-SiC (*21*). $V^{4+}$ has a single electron (spin ½) in its *d* shell ($3d^1$) resulting in a free-ion $^2D$ state that splits into up to five orbital states, depending on the symmetry of the host lattice (*17, 22*). 4H-SiC has two inequivalent sites, one quasi-cubic (*k*) and one quasi-hexagonal (*h*), according to the nearest-neighbor atomic configuration. 6H-SiC has two quasi-cubic sites ($k_1$ and $k_2$) and one quasi-hexagonal site (*h*). A vanadium dopant in a cubic lattice has tetrahedral ($T_d$) symmetry with an orbital doublet ground state ($^2E$) and triplet excited state ($^2T_2$) separated by the crystal field, where the triplet excited state is further split into an orbital singlet ($^2A_1$) and a doublet state ($^2E$) due to spin-orbit interactions, as shown in Fig. 1B. In a hexagonal lattice, the $T_d$ symmetry is further reduced to $C_{3v}$ symmetry due to trigonal distortion that splits all the orbital doublet states. Since the SiC lattice is neither exactly cubic nor hexagonal, one may expect a mixture of these properties. This situation renders exact assignment of the V sites difficult and sometimes contradictory in literature (*16, 23, 24*). For generality, we use the unassigned site names of α and β in 4H-SiC and α, β and γ in 6H-SiC; the question of site assignment will be further discussed along the experimental results. Similarly, the ground and excited states separated by the crystal field are labeled GS1,2 and ES1,2,3 for generality, and are related to the $\Gamma_4$ and $\Gamma_{5,6}$ irreducible representations (*17*).

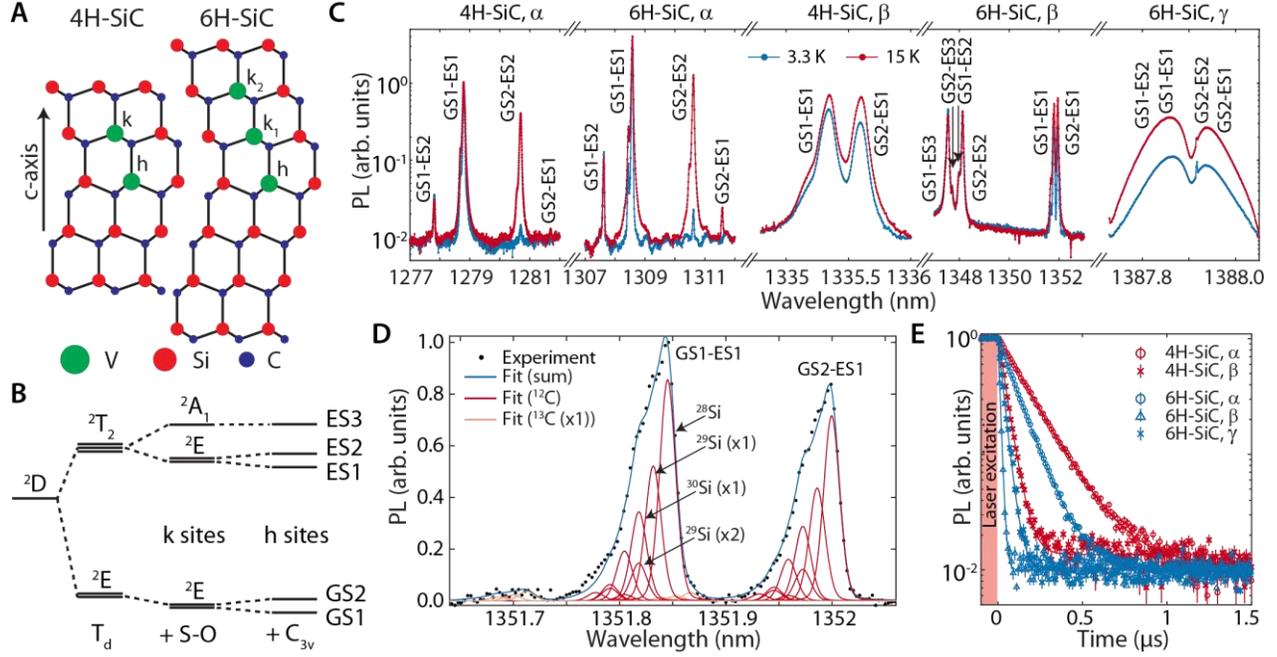

**Fig. 1. Optical spectroscopy of $V^{4+}$ in 4H-SiC and 6H-SiC.** (**A**) SiC lattices and their inequivalent sites for the V impurities. (**B**) Expected energy diagram of the orbital $d^1$ states from crystal field theory and spin-orbit (S-O) coupling according to Ref. (*17, 23*). The quasi-cubic (*k*) sites have mainly a tetrahedral ($T_d$) symmetry and the quasi-hexagonal (*h*) sites have mainly a $C_{3v}$ symmetry due to additional trigonal distortion. (**C**) Resonant PL spectroscopy at 3.3 K (blue) and 15 K (red) for all the available sites. The different transitions are partially identified from the difference in thermal population of the orbital states. The assigned transitions for the 6H-SiC γ site are resolved in Fig. S8, though the weak sharp peak at the center of the spectrum is unknown and is possibly from the laser. (**D**) Resonant PL spectroscopy of the 6H-SiC β site fitted according to an isotope model for the mass shift from neighbor $^{28}$Si, $^{29}$Si, $^{30}$Si, $^{12}$C and $^{13}$C. (**E**) Optical lifetimes at 3.3 K after laser excitation of the GS1-ES1 transition for each available site. All error bars are one standard deviation from experimental acquisition.

We first characterize the $V^{4+}$ orbital structure using resonant excitation of the ensemble optical transitions (Fig. 1C). A tunable laser is set on resonance with the optical transitions, exciting the electronic state and the photoluminescence (PL) emission is collected as the system subsequently decays. The resonant laser is filtered out from the detection by only acquiring the phonon sideband contribution of the PL (see Fig. S1). For 4H-SiC, we stabilize the $V^{4+}$ charge state using a weak above-bandgap illumination (365 nm). Our measurement confirms previous off-resonant studies (*16, 17*) and further resolves the transitions in much finer details. We first observe the similarity between the α sites of 4H-SiC and 6H-SiC by comparing their resonant spectra (and other properties in Table 1). This similarity is likely related to the quasi-identical crystal configuration of the *h* sites in both polytypes (Fig. 1A). The splitting between GS1 and GS2 is largest in the α sites (~500 GHz) resulting in significant thermal polarization at 3.3 K compared to 15 K in Fig. 1C. For the 4H-SiC β and 6H-SiC γ sites, only two transitions are resolved with small ground state splittings of about 10-40 GHz, making it challenging to assign these to specific states. We later solve this issue by looking at the ODMR signal from these transitions (see the Spin properties section) and find that the smallest ground state splitting (16 GHz) belongs to the 6H-SiC γ site, consistent with the most cubic site $k_2$. Finally, only the 6H-SiC β site has six distinct optical transitions that enable the identification of all five orbital states (see Table 1).

For all inequivalent sites, there are consistent features in the optical spectra that were previously unresolved with off-resonant excitation (*17*, *23*). The resonant peaks are asymmetric with a longer tail toward higher energies (lower wavelengths) and are also duplicated at higher energies as shown in Fig. 1D. These duplicates cannot be from additional orbital states which are already fully accounted for in some of the sites. The single-mode behavior of the tunable laser was confirmed with a Fabry-Perot to discard the possibility of an experimental effect. Following previous studies of donor bound excitons in natural and isotopically purified silicon (*25*, *26*), we attribute this asymmetry and duplicates to the presence of the minority isotopes of silicon (4.685% of $^{29}$Si and 3.092% of $^{30}$Si) and carbon (1.07% of $^{13}$C) in the nearest neighbor sites to the vanadium impurity. When a neighbor has a minority isotope, there is a variation in the local mass from the dominant $^{28}$Si and $^{12}$C that shifts the optical transitions, possibly due to an effective local strain or change in the bandgap. For example, a single $^{30}$Si results in twice the shift from a single $^{29}$Si. The shape of the optical peaks is attributed to silicon isotopes, while the duplicate (larger shift) is from the carbon isotopes with larger mass shift ratio than silicon and closer distance to the vanadium.

We model this effect using a multinomial distribution from the natural isotope abundances and nearest neighbor site numbers (see Supplementary Materials S1 and Fig. S2). With only the intrinsic lineshape of the sub-peaks and frequency (wavelength) shift per change in atomic mass as free parameters, we are fully able to reproduce the spectra in Fig. 1D. We find an average shift for all sites of the optical transition frequency by 22(3) GHz/u or $10^{-4}$ %/u for carbon, and by 2.0(5) GHz/u or $10^{-5}$ %/u for silicon. The fit also provides an intrinsic inhomogeneous linewidth of about 2 GHz that may be further reduced considering it includes multiple transitions between the electron and nuclear spin states of ground and excited orbital states. The observed isotope effect suggests that isotopically purified SiC materials may significantly narrow the lines and allow the spin states to be optically resolved. By contrast, quantum registers including both the vanadium nuclear spin and a nearest neighbor carbon or silicon nuclear spin could be directly resolved optically, providing a competitive system for quantum algorithms such as error correction or entanglement distillation. Finally, these shifts hint that the wavefunction of the impurity extends through the first few nearest neighbor shells.

The excited state lifetime is an important parameter to estimate the brightness of a defect. Though previous measurements exist for $V^{4+}$ (*16*), the resonant excitation here excludes possible decay pathways allowed by off-resonant light or possible mixtures of lifetimes from the multiple orbital states. In Fig. 1E, we measure the ensemble excited state lifetime after a pulsed laser excitation between GS1 and ES1. While the α sites have slower decays (108 and 167 ns) than the β and γ sites (11, 31 and 45 ns), the α sites appear brighter. This may be attributed to the α sites being more favorable during the material creation process and therefore more numerous (as seems to be the case with Mo dopants in the hexagonal site of 4H- and 6H-SiC (*15*)), or to higher quantum efficiency in these sites as suggested by density functional calculations in 4H-SiC (*16*). Overall, the lifetimes are much closer to that of vacancy-related defects in diamond and silicon carbide (<

20 ns (*27*, *28*)) than of erbium dopants (~ ms), providing sufficient brightness to observe single V emitters without photonic enhancement. Finally, the Debye-Waller factor for each site is resolved with resonant excitation as shown in Fig. S1, with high values ranging between 25 and 50 % (see Table 1), confirming the viability of vanadium as a telecom emitter of indistinguishable photons.

**Single V centers**

The long lifetime of erbium ions has made observation of single telecom emitters extremely challenging without engineering. Previously this involved either reducing the lifetime via the Purcell effect in an optical cavity (*4*) or by electrical readout via a coupled quantum dot (*11*). Though optical cavities are ultimately necessary for any quantum emitter used for efficient quantum communication, the ability to directly measure single defects using only free-space optics has provided alternative applications (e.g. quantum sensing) and ease of characterization in other systems such as the nitrogen-vacancy center in diamond (*9*). More generally, the stability and reproducibility of single emitters are key challenges to address, notably the requirement of low spectral diffusion of the optical transitions for entanglement applications (*29*, *30*).

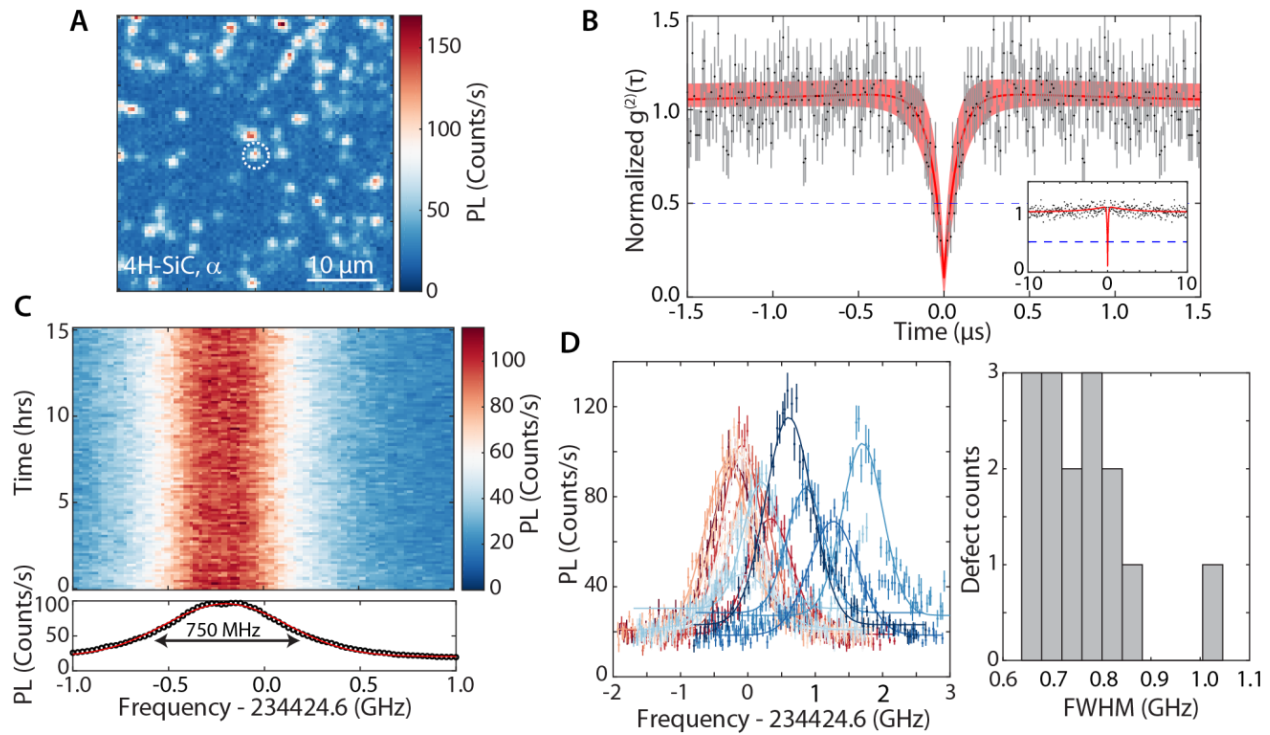

**Fig. 2. Single $V^{4+}$ α site emitters implanted in 4H-SiC.** (**A**) Spatial (near-surface) PL mapping of single and few V defects by resonant excitation at 1278.8 nm and at 3.3 K. (**B**) and (**C**) are obtained at the circled bright spot with spatial feedback to prevent drifting. (**B**) $g^{(2)}$ autocorrelation measurement obtained with a single detector with 20 ns deadtime and 10 ns resolution. The autocorrelation signal is normalized using its value at long delay time and the dark count contribution is calculated and subtracted (~ 3 % of total). The red line is the fit ($g^{(2)}(0) = 0.1(1)$) and the red shadowed area the 95 % confidence interval. In the inset, the autocorrelation intensity is shown for longer times. (**C**) Resonant spectrum taken over 100 acquisitions for a total duration of 15 hours, with averaged intensity shown in bottom. (**D**)

Resonant spectrum taken for a variety of likely single emitters (not confirmed with $g^{(2)}$). Their fitted linewidths (right) remain consistent at about 750 MHz full width half maximum (FWHM). All the single defect experiments are conducted at around 0.1-0.15 T to narrow the linewidth (high field limit). A weak 365 nm continuous illumination helps stabilize the PL from charge conversion (possibly from two-photon ionization). (C) and (D) are calibrated using a wavemeter with below 50 MHz accuracy. All error bars are one standard deviation from experimental acquisition.

We implant $^{51}$V into a 4H-SiC wafer with low defect concentration (no V compensation) and anneal to repair the lattice and incorporate the defects (see Methods and materials section for details). In Fig. 2A, we scan the surface of the sample while exciting the GS1-ES1 transition of the α site. Isolated bright spots corresponding to single or few V defects are observed, with PL intensity around 100-150 counts/s. The combination of low power resonant excitation as well as the high wavelength emission (long-pass filter at 1300 nm) provides a very low background signal limited by the dark counts of the detector (~20 counts/s). We confirm the single characteristic of one of the spots by autocorrelation $g^{(2)}(\tau)$ measurement (Fig. 2B) which shows a clear dip below 0.5 near zero delay ($\tau = 0$). The intensity was obtained using a single photon counter instead of a Hanbury Brown and Twiss configuration and cannot resolve below 20 ns delay, however we obtain a $g^{(2)}(0) = 0.1(1)$ from fit using the relation $g^2(\tau) = 1 - ae^{-\tau/\tau_1} + be^{-\tau/\tau_2}$, where $a$ and $b$ are amplitude parameters and $\tau_1$ and $\tau_2$ are the short and long decay times, respectively. The short decay time $\tau_1$ at 0.07(2) μs is reduced from the 0.17 μs optical lifetime (identical to ensemble) by optical Rabi driving. The autocorrelation also rises slightly above 1 (bunching) farther from $\tau = 0$ and is attributed to a shelving state during optical pumping (with shelving time $\tau_2 \approx 2$ μs) (*31*).

We then characterize the spectral properties of the confirmed single V emitter. The implanted defects are relatively close to the surface with a calculated mean depth of ~100 nm, a configuration that is known to cause strong optical spectral diffusion and blinking for other emitters (*32*). In Fig. 2C, we repeatedly measure the optical spectrum (GS1-ES1 transition) over the course of 15 hours. The spectrum does not shift, jump, broaden or lose intensity during that time and has a linewidth of 750 MHz at full width half maximum. While the linewidth is orders of magnitude larger than the lifetime limit, it is likely broadened by the many unresolved electron and nuclear spin states as measured in the Spin properties section, in addition to the more typical inhomogeneous strain and spectral diffusion. The optical stability is also consistent across various defects with linewidth varying by about ±100 MHz as seen in Fig. 2D. The distribution of peak positions follows the measured inhomogeneous linewidth in ensembles (Fig. 1C). Overall, the optical properties of shallow, implanted single $V^{4+}$ impurities in 4H-SiC appear to be mostly unperturbed compared to ensemble measurements, a crucial result for using these defects in photonic and other monolithic devices. The single defect linewidth remains too broad to resolve the individual spin sublevels, though this may be enabled at higher magnetic fields.

**Spin properties**

$V^{4+}$ in SiC has a single unpaired electron coupled to the $^{51}$V nuclear spin. The spin properties are described by the following Hamiltonian with Zeeman and hyperfine interactions:

$$H = \mu_B \mathbf{B_0} \cdot \mathbf{g} \cdot \mathbf{S} - \mu_N g_N \mathbf{B_0} \cdot \mathbf{I} + \mathbf{S} \cdot \mathbf{A} \cdot \mathbf{I} \quad (1)$$

where $S$ and $I$ are the electron ($S = 1/2$) and nuclear ($I = 7/2$) spin operators, $\mu_B$ and $\mu_N$ are the Bohr and nuclear magneton, $g$ and $g_N$ ($\mu_N g_N = 11.213$ MHz/T) are the electron and nuclear g-factors, $\mathbf{B_0}$ is a static magnetic field and $\mathbf{A}$ is the hyperfine tensor between the electron and nuclear spins. Each of the orbital states therefore split into 16 additional spin states under non-zero magnetic field as illustrated in Fig. 3A. The spin properties of the $V^{4+}$ ground state (GS1) in some sites of 4H-SiC and 6H-SiC were previously studied using electron spin resonance (ESR) and magnetic circular dichroism (*24, 33*), however they were realized at high magnetic field where it becomes challenging to correctly identify anisotropic hyperfine tensors.

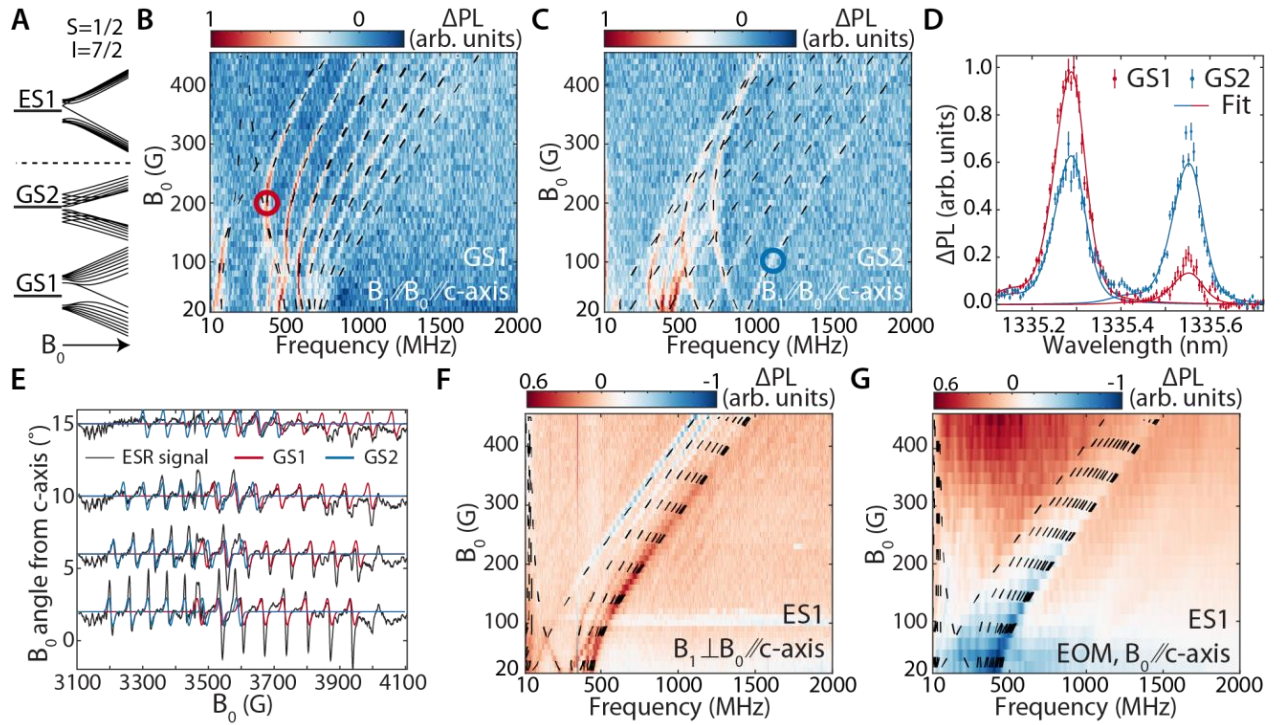

**Fig. 3. Magnetic resonance spectroscopy of the $V^{4+}$ β site in 4H-SiC at 3.4 K.** See Fig. S3-6 for all the other sites in 4H-SiC and 6H-SiC. (**A**) Energy level structure including the electron spin (*S*) and nuclear spin (*I*) for the GS1, GS2 and ES1 levels. The static magnetic field ($B_0$) dependence is simulated using fitted spin parameters from (B), (C) and (F). (**B**) ODMR of the lowest ground state GS1 as a function of $B_0$ and microwave drive frequency. Both $B_0$ and the microwave drive ($B_1$) are applied parallel to the c-axis. The dashed lines in black are modeled from fitting the spin Hamiltonian in Eq. 1. The lines at low magnetic fields around 400 MHz are attributed to ES1. (**C**) ODMR of GS2 under similar conditions to (B). (**D**) ODMR as a function of the laser wavelength to correlate ground states and optical transitions. The red and blue curves are measured respectively at the red and blue circles in (B) and (C). The curves are fitted (using the isotope model lineshape) to extract the individual transitions. (**E**) X-band ESR signal as a function of $B_0$ and angles from the c-axis. The blue and red lines are simulated from the spin parameters obtained by ODMR and are shown to compare derivative peak center positions. (**F**) ODMR of the excited state ES1 with $B_1$ orthogonal to $B_0$. Non-fitted signals in the background are attributed to GS1/2 and are much weaker due to low $g_\perp$ values. (**G**) Optical

hole burning recovery using pump-probe excitations between GS1 and ES1 and detuned by the microwave frequency using an electro-optic modulator (EOM). The transitions are modeled (black dash lines) using the parameters obtained in (F).

We characterize the spin parameters of each site in $V^{4+}$ defect ensembles using a combination of ESR, ODMR and optical pump-probe methods. ESR does not require optical excitation and is consequently sensitive only to the thermally populated ground states. This means that at 3.3 K only GS1 can be measured for the α sites while both GS1 and GS2 may be visible for the β and γ sites. ODMR corresponds to a change in PL intensity when the spin state is rotated by an applied resonant microwave magnetic field ($B_1$). The origin of such contrast is typically due to a shelving state with spin-dependent decay pathways, resulting in different population distribution under steady-state illumination. As ODMR requires optical excitation, it is sensitive to all orbital states with a preference for ground states when the excited states' lifetimes are short. Pump-probe (or hole burning recovery) detection is an all-optical version of ODMR with the microwave field replaced by two optical excitations detuned by the spin frequency, here generated by a single laser and an electro-optic modulator (EOM). This technique is more sensitive to the orbital excited states and more complex depending on the type of pumping scheme (Λ,V,Π,X) (*15*).

We use the 4H-SiC β site to demonstrate the spin measurements as shown in Fig. 3, and provide all the data and fitted parameters for all sites in Fig. S3-6 and in Table 1. The ground states GS1 and GS2 spin parameters are obtained from ODMR (with resonant optical excitation) in Fig. 3B,C by sweeping both the microwave frequency and the static magnetic field $B_0$ (aligned along the c-axis). At low magnetic fields, we observe competition between the hyperfine and Zeeman interactions, enabling precise fitting of the hyperfine tensor diagonal components in their principal axis. We separate the parallel (*zz*) and orthogonal (*xx*, *yy*) components of the electron g-factor by exciting the spins with microwave magnetic fields ($B_1$) either parallel or orthogonal to the c-axis (see Fig. S7). The two ground states GS1 and GS2 can mostly be excited in the parallel configuration, allowed by mixing between the electron and nuclear spins, but only weakly in the orthogonal configuration indicating an orthogonal g-factor $g_{xx,yy}$ close to zero. The hyperfine interactions for GS1 have $A_{xx} = A_{yy} \neq A_{zz}$ (axial symmetry, $d_{z^2}$-like orbital if dipolar (*34*)) for the α sites and $A_{xx} \neq A_{yy} \approx A_{zz}$ (rhombic symmetry, $d_{xz,yz}$-like orbital if dipolar) for the β and γ sites, assuming that equal components have the same sign (not resolved here). These hyperfine aspect ratios may be related to the local symmetry (and corresponding wavefunction distortion), where the Si-C layers stacking is aligned with the c-axis for the *h* sites and tilted for the *k* sites. The GS2 hyperfine tensor was only obtained for the β and γ sites (as thermally populated) and show a non-zero *zz* component with a principal axis tilted by about 50-52° from the c-axis. This angle matches the 52° tilt off the c-axis between two nearest-neighbor silicon sites in separate layers of the SiC lattice.

The assignment of the ODMR signal to an orbital state is not always straightforward as shown in Fig. 3D. Spin resonance ascribed to GS2 (blue line) appears when optically exciting either the GS1-ES1 transition or the GS2-ES1 transition. This indicates that spin polarization is transferred between the two ground states likely from thermal relaxation. The ODMR assignment is resolved in this case by looking at the spin resonance in GS1 (red line) and by comparing the hyperfine tensors of the various sites. The ODMR-resolved resonant spectroscopy of Fig. 3D is also a powerful tool to distinguish close transitions, and allows us to separate the GS1, GS2, ES1 and ES2 orbital states for the 6H-SiC γ site (see Fig. S8).

For consistency, we compare the ODMR signal with ESR experiments in Fig. 3E. Two sets of peaks are observed and assigned to GS1 and GS2 of the 4H-SiC β site only. Similarly, ESR in 6H-SiC presents (Fig. S9) two sets of peaks assigned to GS1 and GS2 of the γ site only. The lack of significant ESR signal from all other sites results from their lower orthogonal g-factor, forbidding microwave transitions at high magnetic fields ($|g\mu_B B_0| \gg |A|$). In general, we find large discrepancies for some of the fitted hyperfine values compared to previous ESR and magnetic circular dichroism experiments (*24*). This is explained by the limited influence of $A_{xx}$ and $A_{yy}$ on the spin resonance frequencies in the high field limit where these previous studies occurred.

**Table 1. Optical and spin properties of V$^{4+}$ defects in 4H-SiC and 6H-SiC around 3.3 K.** The $k_1$ site is assigned to the 6H-SiC β site based on having the closest crystal configuration and properties to the 4H-SiC β site. $k_2$ is the most cubic-like site and therefore assigned to γ with the smallest GS1-GS2 splitting. The Debye-Waller (DW) factor is an upper bound estimation as long wavelength contributions cannot be observed (see Fig. S1). τ is the optical lifetime. The spin parameters (absolute values for the diagonal components $xx$, $yy$, $zz$) are given in their principal axis with $g$ the g-factor and $A$ the hyperfine interaction. $xx$ and $yy$ components are interchangeable, we cannot distinguish $g_{xx}$ from $g_{yy}$, and a good fit is obtained when $A_{xx}$ and $A_{yy}$ are equal for GS2 and ES1. $\theta_{xx}$, $\theta_{yy}$, $\theta_{zz}$ are the angles between the principal and the c-axis basis. "-" indicates unresolved parameters, "*" indicates parameters taken from literature (*23*, *24*) and "**" indicates partially resolved parameters obtained by comparison with other sites.

|  | 4H-SiC | | 6H-SiC | | |
| --- | --- | --- | --- | --- | --- |
| Name | α | β | α | β | γ |
| Site assignment | h | k | h | $k_1$ | $k_2$ |
| ES1 - GS1 (nm) | 1278.808(6) | 1335.331(6) | 1308.592(6) | 1351.845(6) | 1387.806(6) |
| GS2 - GS1 (GHz) | 529(1) | 43(1) | 524(1) | 25(1) | 16(1) |
| ES2 - ES1 (GHz) | 181(1) | - | 167(1) | 628(1) | 6(1) |
| ES3 – ES2 (GHz) | - | - | - | 72(1) | - |
| DW (%) | ≤ 25 | ≤ 50 | ≤ 45 | ≤ 50 | ≤ 40 |
| τ (ns) | 167(1) | 45(1) | 108(1) | 11(1) | 31(1) |
| GS1: $g_{xx,yy}$, $g_{zz}$ | 0*,1.748* | 0<g<1,1.870(5) | 0*,1.749* | -,1.95(2) | 0<g<1,1.933(5) |
| GS1: $A_{xx}$, $A_{yy}$, $A_{zz}$ (MHz) | 165,165,232(5) | 103,188,174(5) | 165,165,232(5) | 114,166,171(5) | 45,215,175(10) |
| GS2: $g_{xx,yy}$, $g_{zz}$ | - | 0<g<1,2.035(5) | - | -,2.00(2) | 0<g<1,1.972(5) |
| GS2: $A_{xx,yy}$, $A_{zz}$ | - | 0,257(5) | - | 0,258(5) | 0,265(5) |
| GS2: $\theta_{xx}$, $\theta_{yy}$, $\theta_{zz}$ (°) | - | 0,52(2),0 | - | 0,50(2),0 | 0,51(2),0 |
| ES1: $g_{xx,yy}$, $g_{zz}$ | -,2.24** | -,2.03(2) | -,2.24* | -,2.0(1) | -,2.03(2) |
| ES1: $A_{xx}$, $A_{yy}$, $A_{zz}$ (MHz) | 20,220(20) | 112,52(5) | 20,200(20) | 80,20(20) | 110,50(20)** |

We finally characterize in Fig. 3F,G the spin properties of the ES1 excited state using both ODMR and pump-probe measurements. The pump-probe optical signal is negative and shows no change when exciting either the GS1-ES1 transition or the GS2-ES1 transition, suggesting the probe is mainly sensitive to the spin levels of the ES1 orbital state. The inhomogeneous broadening in these all-optical measurements is quite large (hole linewidth about 100 MHz, see Fig. S10) and limit precise fitting of the spin parameters. The 4H-SiC β site is unique however as it shows a clear ODMR contrast under orthogonal $B_1$ drive that matches the pump-probe experiment, enabling more sensitive measurement of the ES1 hyperfine parameters. Overall, the ability to obtain the hyperfine and g-factor parameters for three different orbital states offers a unique opportunity to benchmark first-principle calculations of defects in the solid-state. In addition, the large Hilbert space of the electron-nuclear spin system of vanadium offers a rich space for quantum registers, including the use of clock transitions (*19*, *35*) already observed in Fig. 3B (red circle for example).

Coherent control of the spin states as well as long spin lifetimes and coherence times are key figures of merit for quantum information technologies. Probing these parameters is challenging at our lowest available temperature of 3.3 K where the ODMR contrast only starts to become significant for experiments, indicating strong thermal effects. We expect the $V^{4+}$ system undergoes rapid spin relaxation due to phonon processes and the small splitting between the GS1 and GS2 states, similar to silicon vacancies in diamond (*36*). At 3.3 K, we characterize this population relaxation ($T_1$) using hole burning recovery experiment as shown in Fig. 4A,B. A resonant laser pulse excites the GS1-ES1 transition during which the emitted PL intensity decays due to hole burning (population trapping in a shelving state, see Fig. S10). The PL signal requires some delay to regain full intensity after a second laser pulse, corresponding to the thermal relaxation between GS1 and GS2, or to the spin relaxation in GS1, both limiting factors for coherent experiments. The 6H-SiC β site has a short relaxation time of 0.2 μs and the γ site is longest at 1.2 μs, though we are unable to measure the α sites (no significant hole burning). As demonstrated in silicon vacancies in diamond (*36*), these relatively low lifetimes are not necessarily prohibitive as they are expected to significantly increase at millikelvin temperatures.

Finally, we demonstrate in Fig. 4C coherent Rabi driving between two spin states (≈0.74|0.5,−3.5⟩ + 0.65|−0.5,2.5⟩ and −0.67|0.5,−3.5⟩ + 0.74|−0.5,2.5⟩ in |$m_S$,$m_I$⟩ notations) in the GS1 orbital of the 6H-SiC β site. Despite its short lifetime, the corresponding continuous-wave ODMR had a sufficiently narrow inhomogeneous linewidth (about 8 MHz) for driving. We fit the time-dependence by a damped Rabi oscillation formula $\frac{\Omega_R^2}{\Delta\omega^2 + \Omega_R^2 + \Gamma^2} \sin\left(\frac{t}{2}\sqrt{(\Delta\omega+\Gamma)^2 + \Omega_R^2}\right)^2 \exp(-\Gamma t)$ combined with inhomogeneous distribution of drive and detuning, where $\Omega_R$ is the Rabi frequency, $\Gamma$ is a population decay rate and $\Delta\omega$ is a detuning from resonance (*37*). Fig. 4D shows the corresponding pulse-ODMR spectrum after calibrating the π pulse duration from Fig. 4C and confirms the Rabi oscillation is from the spin transition. The decay $\Gamma$ simulates leakage to either GS2 or other nuclear spin states. Over the course of many acquisitions, we observe some decay of the pulse-ODMR contrast, suggestive of a slow nuclear hyperpolarization within the full electron-nuclear spin system (especially to |−0.5,−3.5⟩). The ability to reset the V defect to a specific spin state will be crucial for quantum application and could be achieved either optically or by magnetic resonance (*38*).

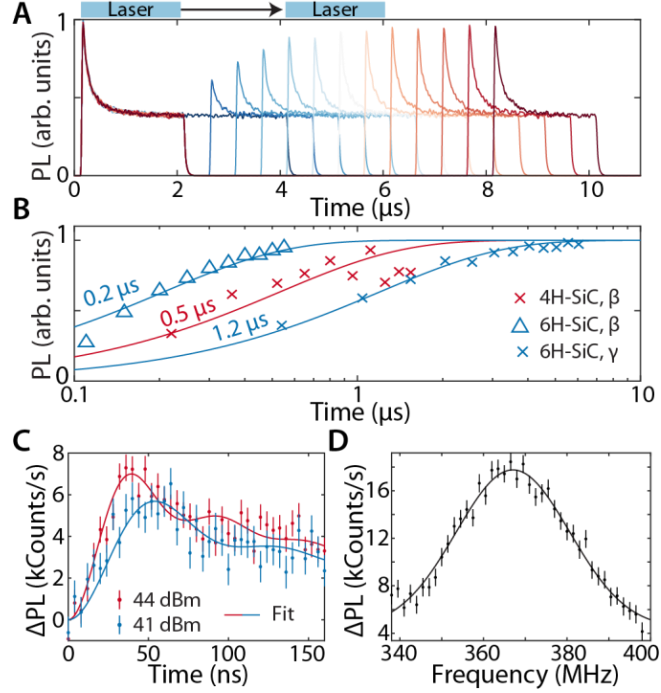

**Fig. 4. Lifetimes and coherent properties at 3.3 K.** (**A**) Transient detection of hole burning and recovery in the 6H-SiC γ site. The transition GS1-ES1 is pumped (0-2 μs) followed by a moving delay before a second pump measures the intensity of the peak signal. (**B**) Optical decays obtained from the sequence in (A). (**C**) Coherent Rabi oscillations obtained in the 6H-SiC β site at 200 G for two different microwave drive powers. Fit is from the equation in main text. (**D**) Pulsed-ODMR spectrum using an inversion pulse calibrated from (C). All error bars are one standard deviation from experimental acquisition.

## Discussion

In this study, we have created and isolated single vanadium centers in commercial SiC with stable spectral properties. This should enable future integration in photonic crystals to enhance the PL intensity by reducing the radiative lifetime (Purcell factor) and improving collection efficiency (fiber coupling). The wavelength shifts by local isotopic impurities observed in ensembles could enable nuclear configuration-selective optical transitions in single emitters, and isotopically purified SiC sample would significantly narrow the inhomogeneous ensemble linewidth. The exact site assignments remain in question, but comparison between the various sites and polytypes suggest the α sites as the quasi-hexagonal site and the $k_1$ and $k_2$ sites in 6H-SiC as the quasi-cubic β and γ sites respectively. Our determination of the g-factor and hyperfine parameters in the various orbital states completes and updates previous experimental results. This work therefore provides all the necessary components for understanding and controlling the $V^{4+}$ defects optically and magnetically, as demonstrated by the coherent Rabi oscillations. Transition metal ions are also known to have strong spin coupling to strain and electric fields (*39*), potentially enabling hybrid spin-mechanical devices. Important questions remain including the spin lifetimes and coherence times at millikelvin temperatures, as well as the exact spin-polarization pathways involved during hole burning. In conclusion, using optical and spin spectroscopy, we have demonstrated the viability of neutral $V^{4+}$ defects in SiC as telecom emitters for quantum applications.

## Materials and Methods
### Samples

Ensemble experiments are realized on V-compensated semi-insulating 4H and 6H-SiC commercial wafers from II-VI Materials. Single defects are created by $^{51}$V implantation with a $10^8$ cm$^{-2}$ dose at 190 keV and 500˚C in a semi-insulating 20 μm epitaxial layer on a N-type 4H-SiC wafer from Norstel. The sample was subsequently annealed at 1400˚C for 30 minutes in Ar to activate the V centers, and with a carbon cap (later removed) to prevent Si evaporation.

### Optical measurements

All optical measurements are realized in a confocal microscopy setup (see Fig. S7) and with the samples mounted in a closed-cycle (Montana Instruments) cryostat with temperatures between 3.3 K and 15 K. Microwave delivery for spin driving uses coplanar waveguides and loop antenna on printed circuit boards placed either beneath or slightly above the sample. All 4H-SiC sites and the α and β sites in 6H-SiC are optically excited using an O-band (1260-1360 nm) tunable laser (EXFO T100S-HP), while the γ site in 6H-SiC is excited using an E-band (1385-1465 nm) tunable laser (Newport Velocity TLB 6700). The laser power from the tunable lasers is about ~100 μW at sample for ensemble measurements, and about ~1 μW maximum for single defects. The laser output is pulsed by an acousto-optic modulator and frequency modulated by a phase electro-optic modulator. The broad spectra are obtained using a liquid nitrogen-cooled InGaAs NIR camera (Princeton Instruments) and are excited off-resonantly by a 365 nm light emitting diode (~μW at sample). The same diode provides charge re-pump/stabilization for all 4H-SiC experiments (*40*). The optical excitation and collection path are set parallel to the c-axis of the sample. The optical illumination is focused on the sample using a 50X Olympus NIR objective and the PL is detected using either a photodiode (FEMTO OE-200-IN1) or a superconducting nanowire single-photon detector (SNSPD, Quantum Opus). SNSPD outputs and timings are directly detected using a 100 MHz data acquisition card. All fit errors are 95% confidence intervals.

### ESR experiments

X-band (9.7 GHz) ESR experiments were undertaken using an ELEXSYS E580 Bruker spectrometer (Bruker Biospin) equipped with a dielectric ring resonator (Bruker EN 4118X-MD4). The samples are mounted in a quartz tube suspended in the center of the resonator, and the ensemble is contained in a flow cryostat (Oxford Instruments CF935) with pumped liquid helium (3.3 K).

**Acknowledgments**

**General:** We thank A. Bourassa, S. J. Whiteley and J. Niklas for experimental assistance and insightful discussions.

**Funding:**
This work is supported by AFOSR, DARPA D18AC00015KK1932, and ONR N00014-17-1-3026. C.P.A. is supported by the Department of Defense through the NDSEG Program, O.G.P. is supported by the U.S. Department of Energy, Office of Science, Office of Basic Energy Sciences, Division of Chemical Sciences, Geosciences and Biosciences, under contract number DE-AC02-06CH11357 at Argonne National Laboratory, and G.W., F.J.H., B.D. and D.D.A. are supported by the U.S. Department of Energy, Office of Science, Office of Basic Energy Sciences. This work made use of shared facilities supported by the UChicago MRSEC (NSF DMR-1420709) and Pritzker Nanofabrication Facility of the PME at the University of Chicago (NSF ECCS-1542205).

**Author contributions:** G.W. conducted and analyzed all the experiments, C.P.A. prepared the samples, B.D. provided critical feedback, G.W., F.J.H. and O.G.P. performed the ESR experiments, and D.D.A. advised on all efforts. All authors contributed to the data analysis and manuscript preparation. The authors declare no competing financial interests. Correspondence and requests for materials should be addressed to D.D.A.

**Competing interests:** The authors declare that they have no competing interests.